# Bootstrap Aggregation for Point-based Generalized Membership Inference Attacks

**Daniel L. Felps[1], Amelia D. Schwickerath[2], Joyce D. Williams,[3] Trung N. Vuong,[4]
Alan Briggs,[5] Matthew Hunt,[6] Evan Sakmar,[7] David D. Saranchak,[8] Tyler Shumaker[9]**

National Geospatial-Intelligence Agency,[1,2,3,4] Concurrent Technologies Corporation,[5,6,7,8,9]
Daniel.L.Felps@nga.mil,[1] Amelia.D.Schwickerath@nga.mil,[2] Joyce.D.Williams@nga.mil,[3] Trung.N.Vuong@nga.mil,[4]
briggsa@ctc.com,[5] huntm@ctc.com,[6] sakmarev@ctc.com,[7] saranchd@ctc.com,[8] shumaket@ctc.com[9]

**Abstract**

An efficient scheme is introduced that extends the generalized membership inference attack to every point in a model's training set. Our approach leverages data partitioning to create variable sized training sets for the reference models. We then train an attack model for every single training example for a reference model configuration based upon output for each individual point. This allows us to quantify the membership inference attack vulnerability of each training data point. Using this approach, we discovered that smaller amounts of reference model training data led to a stronger attack. Furthermore, the reference models do not need to be of the same architecture as the target model, providing additional attack efficiencies. The attack may also be performed by an adversary even when they do not have the complete original data set.

## Introduction

Modern deep neural networks (DNNs) leak information about their training data sets in subtle ways. In the membership inference (MI) attack (Shokri et. al. 2017), an attacker can query the model using a data record and gain an understanding about whether it was used in its training. The vast majority of research into this attack has focused on attack performance and mitigation on a *class resolution*. A newer and potentially stronger attack, the generalized membership inference attack (Long et. al. 2018), shifts the focus to individual points.

In the design of the generalized membership inference attack (GMIA), Long constructed a workflow that permitted the selection of possibly vulnerable points by identifying outlier records, i.e. those that have relatively few neighbors in the embedded space. The motivation was that these are more likely to have a unique influence on the resultant model, a signal that the attack can more readily identify. With this point-based approach, several outlier data points are targeted with high precision when evaluated against a large number of target models. However, Long indicated that identifying all such vulnerable records in the model training set is an open question.

In this work, we introduce an efficient scheme to create a point-based GMIA model for every point in the training set and demonstrate that this creates the strongest known membership inference attack. Our approach leverages data partitioning to create reference models from which samples are used to estimate the target model's population output distribution.

## The Membership Inference Attack

In the MI attack, seemingly benign model queries can be used by an attacker to gain an understanding about which data points were used to train the model. These attacks exploit changes in output behavior observed in models trained by an attacker with and without targeted data points. Numerous attack variants exist depending upon the specific information is available about the model and its output.

Various defensive techniques have been designed for use during model construction to help mitigate this vulnerability. Often these methods are similar to techniques that assist with model generalizability. That is, if the model behaves similarly against new data as it does for training data, then the model is likely both useful and hard to attack and determine the use of specific training data. However, none are perfect and data owners still assume some level of risk in deploying a model.

---

The baseline scenario for practitioners evaluating MI attacks and defenses assumes that the model attacker knows which data set was used for training and that only half of the data set was used. Can the attacker identify which half was used?

In this scenario, random guessing trials would label roughly half as "In" and roughly half as "Out." This results in an area-under-the-ROC-curve (AUC) of 0.5. The vast majority of research into this attack has focused on attack performance and mitigation on a class resolution. As such, attacks are evaluated against each other based upon achieving a better per class and overall AUC score.

**The Class-Based Membership Inference Attack**

In Shokri's MI attack design, an adversary creates "shadow" models that replicate the classification task of a "target" model. The adversary then uses these shadow models to train "attack" models that predict membership. An attack model is trained for every class and they learn membership based upon the class' output probabilities.

The attacker trains the attack models with data sets that they culled, possibly using crowd-sourced or other publicly available information that is readily available and copious in volume. This attack has been used effectively used against both black-box models using commercial machine learning-as-a-Service and white-box models with publicly known structures.

**The Generalized Membership Inference Attack**

Long introduced an attack that shifted the focus to specific training points. In their approach, rather than building one attack model per class, they seek to create one attack model per point. This point-specific attack model relies on the assistance of a large number of reference models.

These reference models were of the same architecture as the target model and were trained using the same amount of data. To accomplish this, they use bootstrap sampling to randomly select a subset of the overall data available to train the target and reference models.

The corresponding output distributions for the target record's classification confidences are used as features in a hypothesis-based attack model.

The reference training data sets were random subsets from a much larger data population, an approach that incorporates the fact that an attacker has access to a significantly larger amount of data not used in the target training. Their approach was also rooted on observing the unique interactions between the target point and other points when one or the other was present and absent.

However, the creation of a large number of reference models is very inefficient in this design. As such, they created a selection mechanism to first help identify what training points could be potentially vulnerable to this attack.

Their assumption is that vulnerable points will likely be outlier records, i.e. those that have relatively few neighbors in the embedded space. The motivation is that these are more likely to have a unique influence on the resultant model, a signal that the attack can more readily identify.

For attack evaluation purposes, they also construct 100 target models, half of which are trained with the target record. They generated training datasets by randomly splitting the records into two datasets of the same size, each serving as a training set for a target model. They repeated this process for 50 times and generated the training datasets for 100 target models.

In their experiments against the MNIST data set, they were able to identify 16 records out of 20,000 whose membership was inferred with greater than 90% precision in over 74% of the target models.

In their related discussions, they highlight that identifying all such vulnerable records in the model training set is an open question. In this work, we introduce an efficient scheme to create a point-based membership inference attack model for every point in the training set and demonstrate that this creates the strongest known membership inference attack.

## Bootstrap Aggregating (Bagging) the Generalized Membership Inference Attack

In this section, we describe an efficient method to construct a point-based attack model for every data point in your population. Our novel contributions come in the construction of the reference models that are used in the training of the attack models.

From the ML practitioner perspective, these attack models can be employed against a model that you seek to deploy in order to understand the most vulnerable points in your training data. However, from the ML attacker perspective, these attack models provide you with a stronger attack that offers confidence about the inclusion of a specific data point in a target model's training set.

Following the approach of Long, we seek to train $k$ reference models, from which we train a point-specific attack model. In their approach, the reference models' training sets are the same size as the target model. The method is depicted in Figure 1.

Our investigation used a modified approach that relaxes the constraint that the reference model's training set be the same size and that the architectures be the same design. Our novel contribution uses data subsampling to efficiently train the large number of reference models.

For each data point in the original training data set, we randomly allocate this point to exactly p of the k reference models. Thus, every reference model has a training data set size of, on average: (overall data set size) * ( p / k).

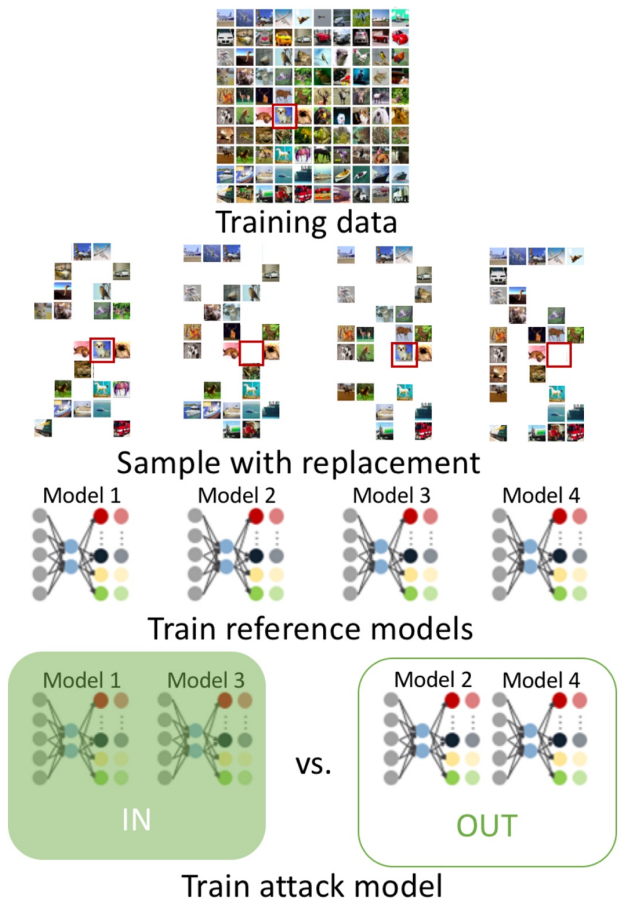

Figure 1: Bagging GMIA with k = 4 and p = 2

for that point. When aggregated, these reference models obtain a representation of the overall training population.

Tables 1 and 2 depict several configurations of these parameters and model architectures used with the CIFAR10 training data size of 50,000. Note that Configuration B is the baseline scenario for the membership inference attack and is intuitively what an attacker most likely to try first – matching the design and training process of the target model. Configuration A can be thought of as an attacker that has seeks advantage via additional attack resources, perhaps using additional data to help understand the interactions encoded in the neural network. Configuration C, D and E represent an attacker with less resources - using only a fraction of the available data.

| Configuration | k | p | Reference Model Training Size |
|---|---|---|---|
| A | 100 | 90 | 45,000 |
| B | 100 | 50 | 25,000 |
| C | 100 | 25 | 12,500 |
| D | 100 | 10 | 5,000 |
| E | 200 | 10 | 2,500 |

Table 1: Example Reference Model Bagging Configurations

| Architecture | Number of Trainable Parameters (millions) | Example Training Time, Configuration B (s) |
|---|---|---|
| Baseline CNN | 0.27 | 720 |
| ResNet20, version 1 | 0.27 | 2514 |
| ResNet56, version 1 | 0.85 | 6180 |

Table 2: Model Architectures and Approximate Train Times

The reference models are trained similarly against the same classification task as the target model. Since any reference model has less data than the target, we generally expect its task performance is to be weaker when compared to deployable model. The extreme case investigated here is when the reference models contain only 5% of the possible training data available.

The use of a large number of reference models can be considered as a statistical sampling of the true output probability for a target model trained with all of the data. As more reference models are used, the more interactions between the data points are incorporated that are statistical estimators of the true interaction. This could also be done using all of the data to create large numbers of reference models of the same size as the target. However, this would require significant resources against large, modern day models.

These reference models are then used to generate inference outputs for each data point that estimate the target model's population output distribution. We then train an attack model for a single point using all refence model output for that point. When aggregated, these reference models obtain a representation of the overall training population.

### Attack Model Construction

The reference model configurations above are trained with the three different architectures, from which inferences (output confidence vectors) are generated for each point in the available data set (i.e., 50,000 for CIFAR10). These inferences will be the features from which an attack model is trained. By default, we do not use data augmentation when training these references models. We will explicitly indicate when reference models have been trained with data augmentation.

The size of each point's feature vector is determined by the number of reference models and the number of output classes. For CIFAR10 and 250 reference models, this vector has size of (250,10). Across all training data points, we have a vector of size (50,000, 250, 10).

Note that the configuration choice of k and p most often results in an imbalanced attack model training data set. That is, a point will be in exactly p models. With 100 reference models and p = 90, this means that the attack model will be trained against 90 confidence features labeled as "In" and 10 confidence features labeled as "Out." Conversely, with 200 reference models and p = 10, there will only be 10 confidence features with the label "In" while there will be 190 labeled as "Out".

In this work, we choose a logistic regression model for ease in both attack confidence interpretability and ability to handle this imbalance. However, a neural network or a statistical test similar to Long's approach would also suffice.

## Experimental Configuration

### Dataset
In this research, we focus on the CIFAR10 dataset. This set is comprised of 50,000 training points across 10 classes.

### Training Environment
All models were constructed in Python 3.6.10 with a TensorFlow 1.13.1 back-end. Model training occurred on a Red Hat Inc. OpenStack Nova VMs configured with Ubuntu 16.04.4 LTS (GNU/Linux 4.4.0-130-generic x86_64). The VMs have access to Tesla V100-PCIe GPUs for model training.

### Target Model Architectures
We train target models using the three different architectures listed in Table 2. However, only parameter configuration B is used in the training. Thus, these models will be trained with 50% of the data and we ensure that each point is in exactly 50 models. In all training, we use a batch size of 128, Adam optimizer, and train for 100 epochs.

**Baseline Convolutional Neural Network**: This model architecture has 0.27M trainable parameters and consists of two sequential blocks of [CONV2 -> RELU -> BN -> CONV2 -> RELU -> BN -> MaxPool], followed by a dense layer with ReLu activation and a Softmax layer. These models are trained without data augmentation.

**ResNet20, version 1** (He et. a. 2015): This model architecture has 0.27M trainable parameters. These models are trained with data augmentation, following common practice for this architecture.

**ResNet56, version 1**: This model architecture has 0.85M trainable parameters. These models are trained with data augmentation, following common practice for this architecture.

### Attack Validation
For attack validation purposes, a set of 100 target models is generated for testing the performance of the above GMIA models and also class-based MI attacks. For each target model, we record the task accuracy and the training time.

For each of the 50,000 possible training points, we perform various attacks against each of the 100 target validation models. Configuration B matches what an attacker most likely attempts first – aligning the attack design with the target model. In addition to Configuration B, we attack with Configuration A and at least one of the remaining Configurations (C through E).

For each point across the 100 target models, we observe and record the following metrics: True Positives (TP), False Negatives (FN), True Negatives (TN), and False Positives (FP). We then compute the number of correct predictions for each point as (TP + TN). For each target model attacked, we calculate the AUC. We average these across the 100 models.

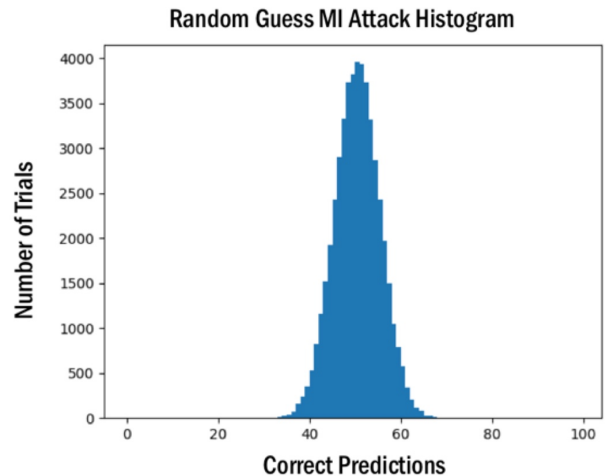

Figure 2: Random Guessing MI Attack Histogram

For an attack based upon 100 random guesses for each of the 50,000 training data points, the expected number of correct predictions across follows a Binomial Distribution with mean 50, standard deviation of 5.0 and an AUC of 0.5. See Figure 2 for a simulated distribution generated in Python. In these results, only 42 training points were attacked with correct accuracy at or above 65%. There were no points attacked with accuracy 75% or above.

In our validation trials, we generate the attack distribution and observe the number of points that can be predicted at or above thresholds of 65%, 75% and 90% accuracy. These points are also compared against those of other target and GMIA architectures to identify points that can be attacked with probabilities higher than expected.

# Results and Analysis

## Baseline Class-Based MI Attack Results

The class-based membership inference attack was employed against each of the 100 target models from each of the architectures. The results of these attacks are in Tables 3 and 4 and Figures 3 and 4.

In both experiments, note that over 30% of the histogram mass is centered on correct attack predictions with 50% effectiveness. Only a small number of points are vulnerable to the class- based membership attack with over 90% accuracy. Compared to the random guess distribution in Figure 2, the shift of mass to greater than 50% accuracy indicates the leakage of training data information.

| Target Model | Mean Accuracy | Mean MI Attack AUC |
|---|---|---|
| ResNet56 | 87.40% | 0.591 |
| ResNet20 | 85.75% | 0.586 |

Table 3: Class-Based Attack Results

| Target Model | 90% | 75% | 65% |
|---|---|---|---|
| ResNet56 | 61 | 3961 | 9974 |
| ResNet20 | 16 | 2494 | 9521 |

Table 4: Class-Based Attack Results

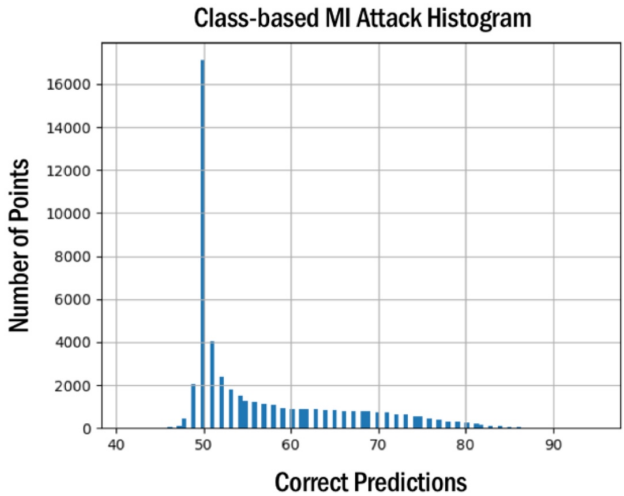

Figure 3: ResNet20 Class Based MI Attack Histogram

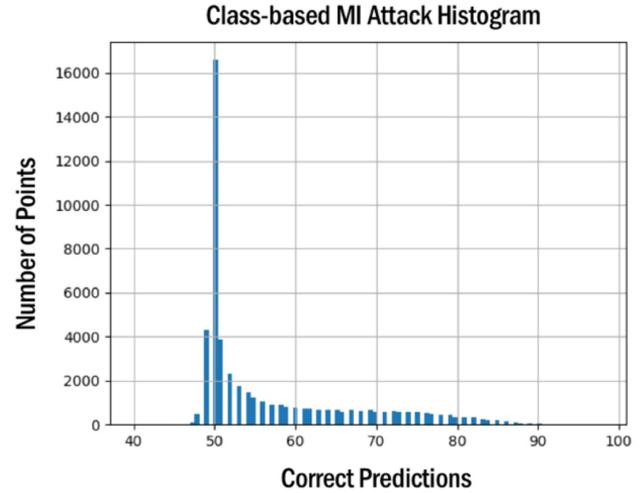

Figure 4: ResNet50 Class Based MI Attack Histogram

## Bagged GMIA Results

The point-based membership inference attack was carried out against the same 100 target models, using different reference model configurations. Results against ResNet56 are listed in Tables 5 and 6, and Figure 5.

| Reference Architecture | Configuration | Mean AUC |
|---|---|---|
| Baseline CNN | A | 0.595 |
| Baseline CNN | B | 0.608 |
| Baseline CNN | E | 0.606 |
| ResNet20 | A | 0.596 |
| ResNet20 | B | 0.612 |
| ResNet20 | D | 0.599 |
| ResNet56 | A | 0.593 |
| ResNet56 | B | 0.614 |
| ResNet56 | E | 0.618 |

Table 5: ResNet56 Bagged GMIA Results

| Reference Architecture | Config | 90% | 75% | 65% |
|---|---|---|---|---|
| Baseline CNN | A | 90 | 2840 | 7647 |
| Baseline CNN | B | 743 | 4954 | 9308 |
| Baseline CNN | E | 377 | 4377 | 8650 |
| ResNet20 | A | 489 | 4686 | 8835 |
| ResNet20 | B | 425 | 4365 | 8680 |
| ResNet20 | D | 403 | 4202 | 8075 |
| ResNet56 | A | 446 | 4620 | 8811 |

Table 6: ResNet56 Bagged GMIA Results. Last 3 columns are the number of training points attacked at or above x% success.

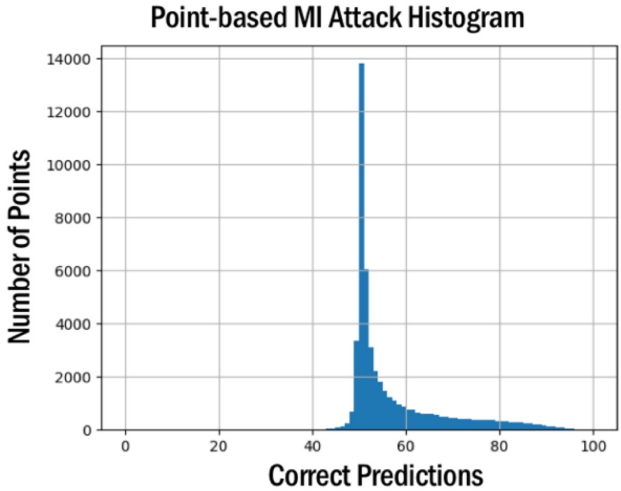

Figure 5: ResNet56 Bagged GMIA Results using ResNet56 Reference Model Configuration E

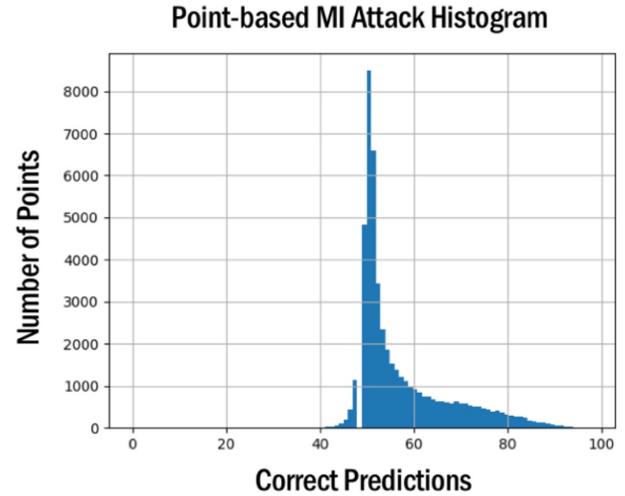

Figure 6: ResNet20 Bagged GMIA Results using ResNet20 Reference Model Configuration E

Results against ResNet20 are listed in Tables 7 and 8, and Figure 6.

| Reference Architecture | Configuration | Mean AUC |
| --- | --- | --- |
| Baseline CNN | A | 0.586 |
| Baseline CNN | B | 0.611 |
| Baseline CNN | D | 0.609 |
| ResNet20 | A | 0.590 |
| ResNet20 | B | 0.615 |
| ResNet20 | E | 0.619 |
| ResNet56 | A | 0.584 |
| ResNet56 | B | 0.618 |
| ResNet56 | D | 0.601 |

Table 7: ResNet20 Bagged GMIA Results

| Reference Architecture | Config | 90% | 75% | 65% |
| --- | --- | --- | --- | --- |
| Baseline CNN | A | 13 | 1528 | 6123 |
| Baseline CNN | B | 160 | 3729 | 9264 |
| Baseline CNN | E | 122 | 3610 | 9328 |
| ResNet20 | A | 148 | 3743 | 8835 |
| ResNet20 | B | 166 | 3776 | 9420 |
| ResNet20 | D | 175 | 3938 | 9586 |
| ResNet56 | A | 128 | 3255 | 8124 |

Table 8: ResNet20 Bagged GMIA Results. Last 3 columns are the number of training points attacked at or above x% success.

### Observations and Analysis

Overall, these results indicate that the Bagged GMIA approach offers both a stronger attack than the class-based approach and an improved way to identify vulnerable training records than Long's GMIA approach.

For the ResNet56 target model, all point-based Bagged GMIA attacks outperformed the class-based method both in terms of mean AUC and the number of vulnerable points identified at all thresholds. Of note is that at the 90% threshold, Baseline CNN Configuration A performed the worst of the Bagged GMIA configurations, but still identified 50% more points than the class-based method. Also, the best Bagged GMIA configuration identified 12 times as many vulnerable points (743 vs. 61).

For the ResNet20 target model, almost all point-based Bagged GMIA attacks outperformed the class-based method both in terms of the mean AUC and the number of vulnerable points identified at all thresholds. Only the ResNet56 Configuration A did not match or outperform the class-based attack for both metrics. At the 90% threshold, the best Bagged GMIA configuration identified 11 times as many vulnerable points (175 vs. 16).

Within the point-based attack configurations that were used to attack ResNet56, it was observed that bootstrap sampling in ResNet56 Configuration E achieved the highest AUC. The Baseline CNN Configuration B performed better than all other configurations in the number of points correctly predicted at the three levels. ResNet 20 reference model architectures performed at levels comparable to ResNet56 reference model configurations.

Within the point-based attack configurations that were used to attack ResNet20, it was observed that bootstrap sampling in ResNet20 Configuration E achieved the highest AUC. Attacking with a more complex model, ResNet56

Configuration D achieved the best performance in terms of number of points correctly predicted at all three percent thresholds. However, using the simpler Baseline CNN and bootstrapping performed comparably well at all three thresholds.

Comparing the specific training images identified via the class-based MI attacks and the GMIA, we observe large intersections. Between the ResNet56 and ResNet20 models attacked with class-based models, they agree on 7,565 points above the 65% threshold, a Jaccard index of 0.634.

For ResNet56 at or above the 65% threshold, there are 8,136 images in common between the class-based attack and the Baseline Configuration E, a Jaccard index of 0.776. There are 8,269 images in common between the class-based attack and ResNet56 Configuration E, a Jaccard index of 0.784. Across the three of these attacks, there are 7,909 images in agreement.

For ResNet20 at or above the 65% threshold, there are 8,052 images in common between the class-based attack and the Baseline Configuration D, a Jaccard index of 0.746. There are 8,073 images in common between the class-based attack and ResNet20 Configuration E, a Jaccard index of 0.732. Across the three of these attacks, there are 7,673 images in agreement.

Comparing the specific training images between the Bagged GMIA of different architectures, we again observe large intersections. For the ResNet56 target model, of the number of points correctly identified above 65%, there are 8,352 in agreement between GMIA ResNet56 Configuration E and the Baseline CNN Configuration E, a Jaccard index of 0.913. At the 90% level, they agree on 340 of the points, a Jaccard index of 0.713. For the ResNet20 target model, of the number of points correctly identified above 65%, there are 8,748 in agreement between GMIA ResNet20 Configuration E and the

ResNet20 Bagged GMIA Results using ResNet56 Reference Model Configuration E Baseline CNN Configuration D, a Jaccard index of 0.861. At the 90% level, they agree on 106 of the points, a Jaccard index of 0.555.

## Discussion and Future Work

In Long's GMIA approach, they were able to identify relatively few training points that were potentially vulnerable in the data sets. For example, in MNIST, they identified 16 out of 20,000 as possibly vulnerable. Their process uses reference models trained with the same amount of data as the target model. They then create new features vectors from pre-softmax layers for vulnerable record selection.

In our Bagged GMIA experiments, we were able to demonstrate an approach that identifies hundreds of training images that are attacked with over 90% certainty. Our method uses reference models with smaller architectures and trained with less data than the target model. Additionally, our method identified thousands of points that are vulnerable across a wide number of models and attacks. As such, this approach is promising in identifying all such vulnerable records in the model training set.

Furthermore, the Bagged GMIA approach offers an efficient way for an adversary to attack a model even if they don't have equivalent resources or the entire training data set, both likely scenarios. Using common hardware to train the three architectures with Configuration B (half of the data), the example train times listed in Table 2 were observed. Using faster GPU hardware decreased these times, but the difference in train times between architectures remained constant. Using these, the time to train 100 Baseline CNN reference models takes roughly 20 hours. In comparison, the time to train 100 ResNet56 models is 171 hours; roughly a factor of 8 longer.

In either case, this "one-time" work can be reused in numerous subsequent attacks and across a wide range of target models. Given a specific point that an adversary wishes to perform bagged GMIA upon against any target model, they can utilize reference models trained with data sets that are vastly different from the true training data set. That is, the approach gives adversaries the ability to model the many possible inference distributions for a point when it is not in the training data set. In comparison, when a data point is included in the training set, its output distribution is generally the same. We interpret these results to be consistent with the Karenina Principle.

## Conclusion

In the Bagged GMIA attack methodology, we have devised an efficient scheme to partition the data and build reference models from which point based membership inference attacks can be performed. We discovered that reference models can be trained faster using smaller amounts of data and this leads to a stronger attack than the class-based attack. Furthermore, we discovered that the reference models do not have to be of the same architecture as the target model, which allows for further resource efficiencies during this attack. Lastly, unlike the original GMIA approach, this works for the entire training set and not just a subset.

## References

He, K.; Zhang, X.; Ren, S.; and Sun, J. 2015. Deep Residual Learning for Image Recognition. arXiv preprint. arXiv: 1512.03385 [cs.CV]. Ithaca, NY: Cornell University Library.

Long, Y.; Bindschaedler, V.; Wang, L.; Bu, D.; Wang X.; Tang, H.; Gunter, C.; and Chen, K. 2018. Understanding Membership Inferences on Well-Generalized Learning Models. arXiv preprint. arXiv:1802.04889 [cs.CR]. Ithaca, NY: Cornell University Library.

Shokri, R.; Stronati, M.; Song, C.; and Shmatikov, V. 2017. Membership inference attacks against machine learning models. *IEEE Symposium on Security and Privacy*, 3-18.